\documentclass[%
twocolumn,
 amsmath,amssymb,
nolongbibliography
]{revtex4-2}

\usepackage{amsmath, amssymb}
\usepackage{comment,xcolor,graphicx}
\usepackage{empheq}

\DeclareMathAlphabet{\mathsfbi}{OT1}{\sfdefault}{bx}{sl}
\DeclareMathAlphabet{\mathsfi}{OT1}{\sfdefault}{m}{sl}
\DeclareMathVersion{sfletters}
\SetSymbolFont{letters}{sfletters}{OML}{ntxsfmi}{b}{it}

\renewcommand{\bm}[1]{\boldsymbol{\mathbf{#1}}}
\renewcommand{\vec}[1]{\bm{#1}}

\providecommand*{\K}{\mathcal{K}}

\newcommand{\A}{\mathcal{A}}
\newcommand{\w}{\mathcal{W}} 
\newcommand{\s}{\mathcal{S}} 
\newcommand{\Rn}{R_n}
\newcommand{\Rt}{R_t}

\newcommand{\ex}{\bm{e}_x}
\newcommand{\ey}{\bm{e}_y}
\newcommand{\ez}{\bm{e}_z}

\begin{document}


\title{
Hydrodynamic Origin of Friction Between Suspended Rough Particles 
}

\author{Jake Minten}
\author{Bhargav Rallabandi}%
 \email{bhargav@engr.ucr.edu}
\affiliation{%
 Department of Mechanical Engineering, University of California, Riverside, California, 92521, USA 
}%





\begin{abstract}
    Tangential interactions between particles play a central role in suspension rheology.  We show theoretically that these interactions, often attributed to contact friction, are a direct consequence of fluid flows between rough particles in relative motion.   We find that small surface asperities generically lead to localized hydrodynamic sliding forces and torques that can exceed their smooth counterparts by orders of magnitude. A fully analytic thin-film theory shows that these forces grow inversely with the surface separation, significantly more singular than the logarithmic scaling for smooth particles. The impending singularity tightly constrains the particles' rotation with their translation, recovering a crucial ingredient in dense suspension rheology. Despite their purely hydrodynamic origin, these features resemble several aspects of  dry rolling and sliding friction. 
\end{abstract}

\maketitle

Particulate suspensions are central to a range of industrial, geophysical, and biological processes. Suspensions of rigid colloids exhibit complex rheological behaviors, including discontinuous shear-thickening (DST), characterized by an orders-of-magnitude increase in viscosity above a critical particle volume fraction \cite{guazzelliRheologyDenseGranular2018,nessPhysicsDenseSuspensions2022}. This is a direct consequence of inter-particle interactions, particularly those that constrain rotational and translational degrees of freedom of the particles \cite{wyartDiscontinuousShearThickening2014, hsiaoTranslationalRotationalDynamics2017, ilhanRoughnessInducedRotational2022}.  However, classical lubrication theory for smooth spherical particles predicts forces that are too weak to enforce these constraints \cite{fossStructureDiffusionRheology2000,melroseContinuousShearThickening2004}. Meanwhile, a growing body of experimental evidence implicates surface roughness as the main cause for particle-scale kinematic constraints and DST \cite{hsiaoRheologicalStateDiagrams2017,ilhanRoughnessInducedRotational2022, pradeepHydrodynamicOriginSuspension2022, hsiaoTranslationalRotationalDynamics2017,schroyenStressContributionsColloidal2019, yanagishimaParticleLevelVisualizationHydrodynamic2021, ilhanRoughnessInducedRotational2022,hsuRoughnessdependentTribologyEffects2018}.
This has led to the notion that particles make physical contact at high volume fractions, marking a transition to (roughness-induced) frictional behavior \cite{setoDiscontinuousShearThickening2013,fernandezMicroscopicMechanismShear2013, wyartDiscontinuousShearThickening2014,mariShearThickeningFrictionless2014, hsiaoRheologicalStateDiagrams2017,townsendFrictionalShearThickening2017}.   

By contrast, Stokesian hydrodynamic simulations \cite{jamaliAlternativeFrictionalModel2019a} that explicitly model the rough particle geometry have recovered rheological signatures without invoking contact friction.  This suggests that lubricated flows between rough particles may be responsible for friction-like interactions. While phenomenological modifications of a lubrication force-law to account for roughness effects have been proposed and utilized \cite{cunhaShearinducedDispersionDilute1996, blancExperimentalSignaturePair2011, wangHydrodynamicModelDiscontinuous2020a}, a systematic theory has been missing \cite{nessPhysicsDenseSuspensions2022}. Theoretical work on rough particle hydrodynamics has focused either on small roughness amplitude or isolated rough surfaces \cite{kurzthalerParticleMotionNearby2020,chaseHydrodynamicallyInducedHelical2022,yarivHydrodynamicInteractionsRough2024}, and thus do not explain the necessary rotational constraints. 

We develop a rigorous thin-film theory for interactions of suspended rough particles close to contact. We show that hydrodynamic flows at the scale of roughness asperities lead to large point-like tangential forces and torques that greatly exceed their smooth counterparts. The forces grow as the inverse of the separation distance $1/d$, in contrast with the weak $\log d$ scaling for smooth spheres. These hydrodynamic singularities enforce kinematic constraints on particle rotation even in the absence of physical contact, yet exhibit characteristics reminiscent of dry friction.  

\begin{figure*}[t!]
    \centering
    \includegraphics[width=0.9\textwidth]{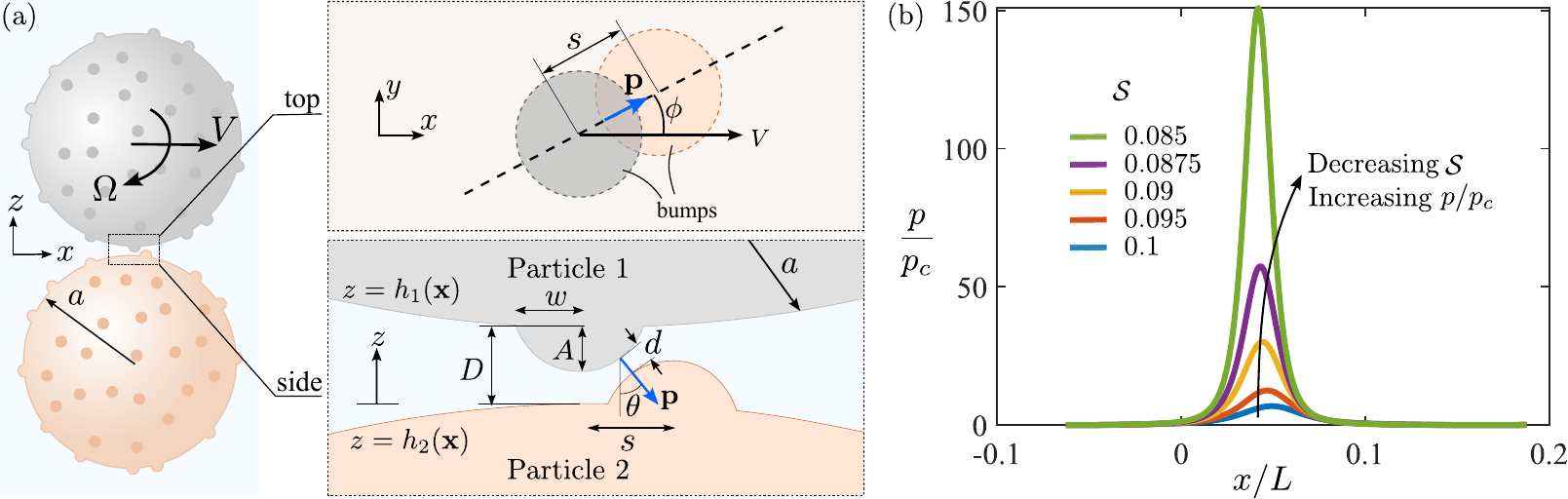}
    \caption{(a) Two spherical particles of radius $a$, with asperities of width $w$ and amplitude $A$, rotate and translate relative to each other. 
    The microscale separation $d$ between asperities can be much smaller than the nominal separation $D$ between the particles, leading to singular hydrodynamic forces in the approach to contact. (b) Pressure profile in the gap at  $y=0$, showing a sharp spike as the bumps approach contact (decreasing $\s$).} 
    \label{FigSetup}
\end{figure*}

We consider the near-contact hydrodynamics of two spherical particles of radius $a$, whose surfaces are decorated with small roughness asperities (bumps); see Fig.~\ref{FigSetup}a. The particle surfaces are separated by a nominal ``macroscopic'' gap $D$, while the ``microscopic'' gap $d$ -- defined as the minimum separation between the asperities -- can be much smaller. Hydrodynamic features are highly localized to the small gaps between opposing asperities, so we focus on a single pair of asperities (one per particle). The bumps have amplitude $A$, radial width $w$, a relative orientation angle $\phi$, and a center-to-center distance $s$ in the plane of the gap. The amplitude is small relative to the particle radius, but can be comparable to the gap $D$.

We allow the top particle (particle 1) to translate with velocity $V \bm{e}_x $ and rotate with angular velocity $\Omega \bm{e}_y $, while keeping the lower particle (particle 2) fixed; more general kinematics will be treated elsewhere. We consider small gaps and gentle bump profiles, i.e. $D \ll a$ and $A \ll w$, providing the necessary conditions for lubrication theory. The generic curved geometry near the point of nearest separation is paraboloidal. The rough particle surfaces are encoded by $z = h_{1, 2}(\bm{x})$, where $\bm{x} = (x,y)$ represents locations in the plane of the gap; see Fig. \ref{FigSetup}(a) and Supplemental Material \cite{SI} for details. 

Rather than follow the trajectory of the asperities through time, we  characterize the system in configuration space. Defining the gap profile $h(\bm{x}) = h_1 - h_2$ and twice the mean gap surface $\overline{h}(\bm{x})=h_1+h_2$, the pressure $p(\bm{x})$ is governed by the Reynolds equation \cite{ockendon1995viscous,SI}
\begin{align}
    \nabla&\cdot\left[\frac{h^3}{12\mu}\nabla p+\frac{\overline{h}}{2}\left(V-a \Omega \right)\vec{e}_x\right]  +\Omega x=0, 
    \label{eq:ReynoldsLubrication}
\end{align}
where $\mu$ is the viscosity of the fluid and $\nabla$ is the gradient in the $xy$ plane. We non-dimensionalize by scales characteristic of the smooth limit: distances across the gap by $D$, horizontal distances by $L = \sqrt{2 a D}$, and the pressure by  $p_c = \mu VL/D^2$. The rescaled problem involves the dimensionless amplitude $\A = A/D$ and width $\w = w/L$ of the asperities, their relative in-plane orientation $\phi$ and separation $\s = s/L$, which together set the dimensionless gap $\delta = d/D$.

We numerically solve the rescaled form of \eqref{eq:ReynoldsLubrication} for $p(\bm{x})$, \cite[][pp. 1066]{press2007numerical}, and then integrate the stresses of the flow to obtain forces and torques exerted by the flow on the particles (see \cite{SI} for details). To aid the discussion, we temporarily suppress rotation and focus on bumps oriented head-on ($\phi = 0$); both constraints will be relaxed later. 

We first discuss the horizontal force $F_x$, which we decompose into a combination of smooth $ F_{x,S}$ and rough  $F_{x,R} = F_x - F_{x,S}$ contributions.  Including Stokes drag, the smooth contribution is $F_{x, S} \approx 6\pi F_c \left[\frac{1}{6}\log \left(\frac{a}{D}\right) + 1\right]$, where $F_c = \mu a V$ is the characteristic force scale on the particle \cite{jeffreyCalculationResistanceMobility1984}. The logarithmic term is due to smooth-particle lubrication and depends only weakly on the gap (i.e. it is only about as large as Stokes drag even for $D/a$ as small as $10^{-3}$).

The rough contribution to the force, which is our focus, depends on the inter-bump separation $\delta$, which we control by varying the location $\s$ of the bump on particle 2. The bump on particle 1 is centered for simplicity, but this is nonessential for our discussion.  For rough spheres, we find that the pressure is much greater than the scale expected for smooth spheres, and shows a highly localized spike that grows as bumps are brought closer together, which decreases $\delta$ (Fig.~\ref{FigSetup}b). 

\begin{figure}[h!]
    \centering
    \includegraphics[width=0.35\textwidth]{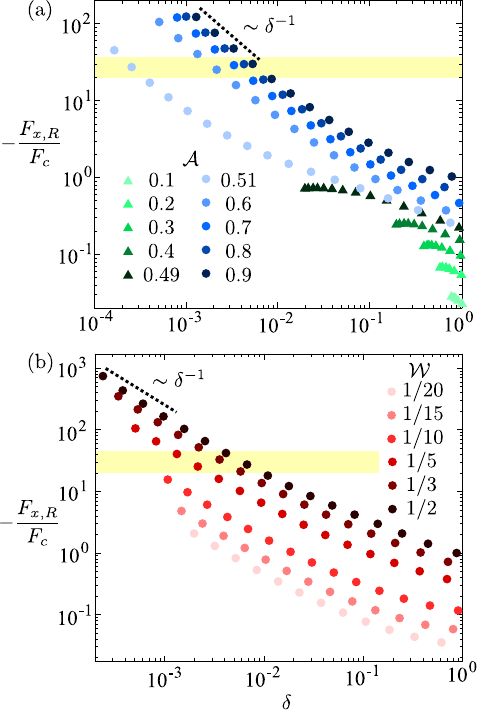}
    \caption{Dimensionless horizontal force (rough contribution) versus gap height $\delta$ for a pair of opposing asperities. The force $F_{x, R}$ scales as $\delta^{-1}$ for $\A > 1/2$, and exceeds the smooth contribution (shaded region). Results at (a) fixed $\w=1/5$ for a range of $\A$ (b) fixed $\A=0.6$ for a range of $\w$. The force increases with both $\A$ and $\w$. } 
    \label{FigNumerics}
\end{figure}

These pressure spikes correlate with increased \emph{horizontal} forces, unlike smooth particles where the horizontal force is largely caused by shear. Increasing either the amplitude or width of the asperities increases $F_{x,R}$ (Fig. \ref{FigNumerics}). When $\A < 0.5$, all horizontal positions $\s$ are accessible, and the rough contribution to the force remains negligible relative to the smooth case (Fig. \ref{FigNumerics}a). This behavior changes qualitatively as soon as $\A$ exceeds $0.5$. Now, the asperities are in a state of \emph{impending} contact under a tangential sliding of the particles, restricting the available configuration space $\s$ (cf. Fig. \ref{FigSetup}a). In this regime, $\delta$ approaches zero as the bumps come closer, leading to a sharp increase in force. Even for a single pair of asperities, this rough hydrodynamic force overwhelms the smooth-sphere sliding lubrication forces (shaded regions in Fig. \ref{FigNumerics}). The numerical results suggest that the force diverges as a power-law $\delta^{-1}$, which lies in stark contrast with the much weaker (logarithmic) smooth contribution. This scaling is consistent with a squeeze-flow between the asperities due to tangential sliding of the particles, suggested in previous work \cite{jamaliAlternativeFrictionalModel2019a,wangHydrodynamicModelDiscontinuous2020a}.

To quantify these features we zoom into a ``local'' region between the asperities (see the side view of Fig. \ref{FigSetup}a). In this region, the asperities appear as spheres of radius $b= a \kappa/(1 + \kappa)$, separated by a gap $d$, where $\kappa = w^2/(2Aa) = \w^2/\A$ is a measure of the bump's curvature relative to that of the particle. The hydrodynamics of the inter-bump gap are governed by a horizontal length scale $\ell = \sqrt{2 b d} \propto w\sqrt{d/A}$, which becomes much smaller than both $w$ and $L$ as the bumps approach contact. We exploit this separation of scales to isolate the bumps from the rest of the geometry, and invoke lubrication theory once more, this time in the thin film between the two spherical bump surfaces in relative translation. The force on bump 1 in the local configuration takes the form
\begin{align} 
    \bm{F}=-\mu b \left[\Rn\bm{p}\bm{p}+\Rt\left(\bm{I}-\bm{p}\bm{p}\right)\right]\cdot (V \bm{e}_x),
    \label{eq:F_x_vectortheory}
\end{align}
where $\Rn$ and $\Rt$ are dimensionless resistance coefficients for normal and tangential motion \emph{between bumps}, respectively, and $\bm{p} = \sin\theta\cos\phi~\ex+\sin\theta\sin\phi~\ey-\cos\theta~\ez$ is the unit vector which connects the centers of the two bumps and is normal to their surfaces (Fig. \ref{FigSetup}a). Also, $\phi$ is the orientation angle introduced earlier, while $\theta$ is a polar angle that depends on the relative position of the bumps. 

When the distance between asperities is small ($d \ll b$), the  resistances are
\begin{align} \label{Resistances}
    \Rn=\frac{3\pi}{2} \frac{b}{d}, \quad 
    \Rt=\frac{\pi}{2}\log\left(1+\frac{A}{d}\right).
\end{align}
The normal resistance $\Rn$ is a well-known result \cite{cox_suspended_particles_1974}, whereas the tangential resistance $\Rt$ is slightly modified from the usual result as the outer cutoff of the logarithm is set by $w$ rather than $b$; see \cite{SI}. Importantly, $\Rn$ scales like $d^{-1}$ and is associated with a squeeze flow between bumps, while $\Rt$ grows weakly as  $\log d^{-1}$. 

Substituting \eqref{Resistances} into \eqref{eq:F_x_vectortheory} leads to general expressions for the force components
\begin{subequations} \label{FLocalphi}
 \begin{align}
    F_{x,R}&=-\mu b V\left[(\Rn-\Rt)\sin^2\theta\cos^2\phi+\Rt\right],
    \label{eq:F_x,R} \\
    F_{y,R}&=-\mu b V(\Rn-\Rt)\sin^2\theta\sin\phi\cos\phi,
    \label{eq:F_y,R} \\
    F_{z,R}&=\mu b V(\Rn-\Rt)\sin\theta\cos\theta\cos\phi.
    \label{eq:F_z,R}
 \end{align}
\end{subequations}
The polar angle $\theta$ is a function of the microscopic geometry ($\A$, $\w$, and $\s$) and thus depends implicitly on $\delta$. Under a small-angle approximation consistent with the thin-film geometry, we find $\sin\theta\approx\sqrt{\frac{(2A-D+d)(1+\kappa)}{a \kappa (1+2\kappa) }}$ \cite{SI}. Substituting this result into \eqref{eq:F_x,R} yields 
\begin{align}
    F_{x,R}= -\mu a V \K \bigg[\frac{3\pi}{2} \left(\frac{2 \A - 1}{\delta} + 1\!\right)\!\cos^2\phi \! +\Rt\bigg],
    \label{eq:F_x,R small}
\end{align}
where $\mathcal{K} =  \kappa/[(1+\kappa)(1 + 2 \kappa)] = (b/a) (1 - b/a)/(1 + b/a)$ depends on the ratio of the bump to the particle curvatures; note that  $\K \sim \kappa = w^2/(2 A a)$ for small $\kappa$.  We find the tangential rough contribution $\Rt$ to be much smaller than even the smooth case \cite{SI}, so we neglect it for the remainder of this article.  

The analytic theory \eqref{eq:F_x,R small} fully reproduces the features of the numerical results in Fig. \ref{FigNumerics}. When $\A < 1/2$, the bumps can pass over each other, so the more singular $\Rn$ contribution vanishes at closest approach ($\theta = 0$, where $\delta = 1 - 2\A$) and only the smaller logarithmic $\Rt$ contribution survives. By contrast, for taller bumps ($\A>1/2$), the squeeze flow between asperities generates a $\delta^{-1}$ hydrodynamic force that resists contact. Despite its small prefactor, the $\delta^{-1}$ singularity from a single bump-pair quickly overwhelms the combined ``smooth'' Stokes drag and lubrication forces on the rest of the sphere (shaded region in Fig. \ref{FigNumerics}).

\begin{figure}[t!]
    \centering
    \includegraphics[width=0.43\textwidth]{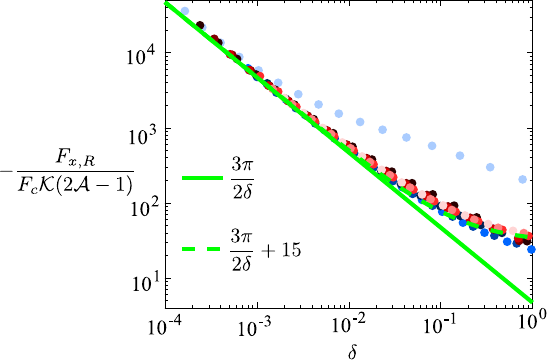}
    \caption{Horizontal force $F_{x,R}$ rescaled by the theoretically predicted scale $F_c \mathcal{K} (2 \A - 1)$, plotted against $\delta$. Symbols are data from Fig. \ref{FigNumerics}, which collapse into the theoretical prediction $3\pi/2\delta$ for small $\delta$ (solid line) for all $\A$ and $\w$, without fitting parameters. The collapse persists for larger $\delta$, and is captured by adding a subdominant empirical constant $c = 15$ to the theoretical prediction (dashed curve).} 
    \label{FigCollapse}
\end{figure}

We quantitatively test the theory in the singular regime $\A > 1/2$. As the bumps approach contact, \eqref{eq:F_x,R small} reduces to
\begin{align}
    F_{x,R} \sim - F_c \frac{3\pi \K}{2} \frac{(2\A-1)}{\delta}  \cos^2\phi, \quad  \A>1/2. 
    \label{eq:F_x,R approx}
\end{align}
We use this result to rescale the numerical results of Fig. \ref{FigNumerics} by the scaling factor $(2 \A - 1) \K $. The rescaled data collapse onto the theoretical prediction $3\pi/2\delta$ for small $\delta$, without adjustable parameters (Fig~\ref{FigCollapse}). Unexpectedly, this  collapse persists even when $\delta$ is not small, and is described rather well by a universal curve $-3 \pi/2 \delta + c$, where we find the subdominant term $c \approx 15$ from a fit.  

We thus see that the rough contribution is essentially ``switched on''  when $A > D/2$ ($\A  > 1/2$), which corresponds to \emph{impending contact} between asperities under a sliding of the particles. This threshold is observed clearly in the numerical results, and emerges naturally in the theory. In a situation with multiple bumps, the threshold $A > D/2$ corresponds to configurations where particles can interlock, which has been identified as important to DST in past work \cite{hsuRoughnessdependentTribologyEffects2018}. However, while the interactions of interlocking particles were thought to be driven by contact-friction,  we see that they are in fact a direct consequence of the hydrodynamics of the approach to contact. This can only occur when both surfaces are rough; if one of the surfaces is smooth, only the subdominant $\Rt$ term survives (cf. \cite{yariv_rough_surfaces_2024} for a two-dimensional analog).
 
The theory provides a natural framework for arbitrary orientations $\phi$ between asperities. The predicted $\phi$ dependences in \eqref{FLocalphi} are confirmed by our numerical solutions (see \cite{SI}).  At fixed $\phi$, all three force components depend on $\Rn$, diverging as $\delta^{-1}$ for $\A  > 1/2$.  However, not all three components are equally important \emph{on average}. Any physically realizable system of rough particles will involve an ensemble of pairwise bump interactions with different orientations at any instant of time. If bumps are distributed isotropically on the particles, all orientations $\phi \in [0, 2\pi)$ are realized with equal probability. Then, averaging \eqref{FLocalphi} over  $\phi$ leads to the average force per pair of opposing bumps,
\begin{align}     \label{eq:F_x,R avg} 
    \left<F_{x,R}\right>=- F_c \frac{3\pi \K}{4} \frac{(2\A-1)}{\delta}, \; \left<F_{y,R}\right> =  \left<F_{z,R}\right>=0,
\end{align}
valid for $\A > 1/2$. Only forces in the direction of motion ($x$) survive the averaging since $F_x$ maintains its sign regardless of orientation,  while the other components average out to zero, due to orientational symmetry at the asperity scale. Thus, the configurational averaging retrieves the macroscopic Stokesian symmetry of the two-sphere system.


We now generalize to include both rotation $\Omega \bm{e}_y $ and translation $V \bm{e}_x$, focusing on how these degrees of freedom are coupled by roughness. We also account for the torque on the particle, only the   $y$-component of which is relevant in the average sense. For the same reason, we neglect vertical translation, which will not, on average, couple to horizontal translation or rotation due to symmetry. Under this mean-field setting, the mobility relation connecting a tangential force and torque due to tangential sliding and rotation of particle 1 is 
\begin{align}
    \begin{bmatrix}
         F_x \\
         T_y \\
    \end{bmatrix}
    = -\mu a \begin{bmatrix}
        \mathcal{R}_{F V} & a \mathcal{R}_{F \Omega}  \\
        a \mathcal{R}_{T V} & a^2 \mathcal{R}_{T \Omega}  \\
    \end{bmatrix} \cdot
       \begin{bmatrix}
         V \\
         \Omega \\
    \end{bmatrix},
    \label{ResistMatrix}
\end{align}
where the $\mathcal{R}_{ij}$ are dimensionless resistance coefficients (note that $\mathcal{R}_{F\Omega} =  \mathcal{R}_{TV}$ due to symmetry). 

The $\mathcal{R}_{FV}$ coefficient relates the force to the particle velocity. By definition, it is identical to the quantity $F_{x,R}/F_c$ in Figs. \ref{FigNumerics} and \ref{FigCollapse}, where we had set $\Omega = 0$. To understand (the rough contributions to) the other coefficients we revisit the local theory. Including particle rotation merely modifies the approach velocity of the asperities in \eqref{eq:F_x_vectortheory} from $V \bm{e}_x $ to $(V - a \Omega) \bm{e}_x $, which carries through to all subsequent theoretical results for the force. Then, the rough contribution to the torque on particle 1 is 
    ${\bm{T}_R = \int \bm{r} \times \bm{n} \cdot \bm{\sigma}_R  dS}$, 
where $\bm{\sigma}_R$ is the stress due to roughness and $\bm{r}$ is the moment arm connecting the center of the particle to a point on its surface. As shown in Fig. \ref{FigSetup}(b) and predicted by the theory, stresses are localized to a small region of length $\ell \ll L$ surrounding the minimum separation of the bumps. Within this region, the moment arm is effectively the constant vector $\vec{r}_c$ connecting the center of particle 1 to the point of minimum separation. The torque is therefore $\bm{T}_R \approx \bm{r}_c \times \bm{F}_R$, where $\bm{F}_R = \int\bm{n} \cdot \bm{\sigma}_R dS$ is the rough contribution to the force discussed previously. This cross product does not vanish even though $\bm{F}_R$ is largely a normal force (directed along $\bm{p}$) since $\bm{p}$ is misaligned with the moment arm $\bm{r}_c$  (this is not the case for smooth spheres). Then, the torque on the particle is simply a consequence of the rough force acting at an effective point of contact. We thus conclude that all four resistance coefficients are identical up to sign, 
\begin{align} \label{ResistTheory}
    \mathcal{R}_{FV} &= -\mathcal{R}_{F\Omega} =  -\mathcal{R}_{TV} =  \mathcal{R}_{T\Omega} \nonumber \\ 
    &= \frac{3 \pi \K}{2} \frac{(2 \A - 1)}{\delta} \cos^2\phi, \quad (\A > 1/2).
\end{align}
\begin{figure}[t!]
    \centering
    \includegraphics[width=0.48\textwidth]{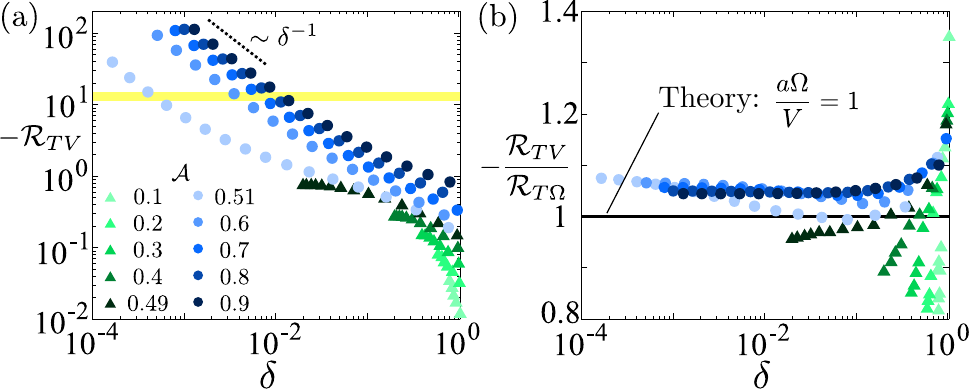}
    \caption{(a) Resistance coefficient $\mathcal{R}_{TV}$ (showing rough contributions only) versus $\delta$, for fixed $\w=1/5$ and various $\A$. The data are virtually identical to those in Fig. \ref{FigNumerics}a. Shaded region indicates a typical range for smooth particles for $D/a \in (10^{-3}, 10^{-1})$. 
    (b) Ratio of resistance coefficients $\mathcal{R}_{TV}/\mathcal{R}_{T \Omega} = a \Omega/V$ versus $\delta$. For all $\A$ and all $\delta$, the ratio is always close to 1, indicating strong rotation-translation coupling.}
    \label{fig:Torque}
\end{figure}

Our numerical calculations quantitatively corroborate these predictions. The torque due to translation, $\mathcal{R}_{TV}$, is virtually identical to $\mathcal{R}_{FV}$ (compare Fig. \ref{fig:Torque}(a) with Fig. \ref{FigNumerics}(a)), diverging like $\delta^{-1}$ for $\A>1/2$ and exceeding the torque on a smooth particle. If particle 1 is torque-free, its rotation and translation are coupled according to
\begin{align}
    \frac{\Omega a}{V}=-\frac{\mathcal{R}_{T V}}{\mathcal{R}_{T \Omega}}. 
\end{align}
For small gaps, the torques from roughness dominate those from the rest of the two-particle geometry, and this ratio approaches unity according to the theory \eqref{ResistTheory}, corresponding to $\Omega a/V \approx 1$. This prediction is in agreement with the numerical results (Fig. \ref{fig:Torque}b), and corresponds precisely to vanishing relative surface velocity $V - a \Omega$. This ``no-slip'' condition is established to relieve impending $d^{-1}$ rough-hydrodynamic singularity, recovering the constraint of classical dry rolling friction.  We contrast this with the case of smooth particles, where the coupling is negligible \cite{cox_suspended_particles_1974,hsiaoTranslationalRotationalDynamics2017}, i.e. torque-free smooth particles experience a very weak coupling that decays rapidly with $D$. The roughness-induced torque dominates over the smooth contribution close to contact (i.e., when $A > D/2$), and so we expect rotation and translation to be strongly coupled in this regime. 
This coupling is insensitive to the geometric details of the roughness, and stems from the generation of large forces in the ``interlocking regime'', as well as the localization of the stresses on sub-asperity length scales. Such a coupling places strong constraints on the kinematics of dense suspensions, and is central to understanding DST. 

%
We have shown that local hydrodynamic interactions between asperities generically lead to a strong singular behavior of tangential forces and torques on rough particles near contact. Even for a single pair of bumps, these forces can exceed hydrodynamic forces between smooth particles, and occur when the nominal separation between particles falls below twice the roughness amplitude. On the scale of the particles, the roughness-induced hydrodynamic features manifest as localized forces that act at an effective point of contact. These point-like forces generate torque, which also greatly exceeds its smooth counterpart. These large forces and torques tightly constrain rotation to translation.

On the one hand, the forces discussed here share many similarities with frictional contact forces: they are point-like, and occur below a threshold separation on the order of the roughness amplitude. Macroscopically, this threshold can be viewed as an effective contact distance at which these forces become ``activated.'' On the other hand, they are purely hydrodynamic and scale with particle velocity, and become singular as the inverse of the \emph{microscopic} separation distance $d$. These hydrodynamic forces and torques occur precisely to \emph{avoid} an impending hard contact between asperities. Thus, rough hydrodynamic forces form a bridge between smooth hydrodynamic interactions and frictional contact forces. 

Our findings form a fundamental basis for near-contact interactions in suspensions of rough particles. Roughness produces a singular perturbation to lubrication theory, leading to large hydrodynamic resistances in near-contact configurations. These modified resistances would be important --- and relatively straightforward --- to include in simulations of dense suspension flows. The same principles would be broadly applicable to other systems involving moving surfaces close to contact. \\ 

The authors thank Vincent Bertin, Howard Stone, Alex Greaney,  and Ehud Yariv for insightful conversations. JM acknowledges support from the US Department of Education through award P200A210080. 


\begin{thebibliography}{29}%
\makeatletter
\providecommand \@ifxundefined [1]{%
 \@ifx{#1\undefined}
}%
\providecommand \@ifnum [1]{%
 \ifnum #1\expandafter \@firstoftwo
 \else \expandafter \@secondoftwo
 \fi
}%
\providecommand \@ifx [1]{%
 \ifx #1\expandafter \@firstoftwo
 \else \expandafter \@secondoftwo
 \fi
}%
\providecommand \natexlab [1]{#1}%
\providecommand \enquote  [1]{``#1''}%
\providecommand \bibnamefont  [1]{#1}%
\providecommand \bibfnamefont [1]{#1}%
\providecommand \citenamefont [1]{#1}%
\providecommand \href@noop [0]{\@secondoftwo}%
\providecommand \href [0]{\begingroup \@sanitize@url \@href}%
\providecommand \@href[1]{\@@startlink{#1}\@@href}%
\providecommand \@@href[1]{\endgroup#1\@@endlink}%
\providecommand \@sanitize@url [0]{\catcode `\\12\catcode `\$12\catcode
  `\&12\catcode `\#12\catcode `\^12\catcode `\_12\catcode `\%12\relax}%
\providecommand \@@startlink[1]{}%
\providecommand \@@endlink[0]{}%
\providecommand \url  [0]{\begingroup\@sanitize@url \@url }%
\providecommand \@url [1]{\endgroup\@href {#1}{\urlprefix }}%
\providecommand \urlprefix  [0]{URL }%
\providecommand \Eprint [0]{\href }%
\providecommand \doibase [0]{https://doi.org/}%
\providecommand \selectlanguage [0]{\@gobble}%
\providecommand \bibinfo  [0]{\@secondoftwo}%
\providecommand \bibfield  [0]{\@secondoftwo}%
\providecommand \translation [1]{[#1]}%
\providecommand \BibitemOpen [0]{}%
\providecommand \bibitemStop [0]{}%
\providecommand \bibitemNoStop [0]{.\EOS\space}%
\providecommand \EOS [0]{\spacefactor3000\relax}%
\providecommand \BibitemShut  [1]{\csname bibitem#1\endcsname}%
\let\auto@bib@innerbib\@empty
\bibitem [{\citenamefont {Guazzelli}\ and\ \citenamefont
  {Pouliquen}(2018)}]{guazzelliRheologyDenseGranular2018}%
  \BibitemOpen
  \bibfield  {author} {\bibinfo {author} {\bibfnamefont {{\'E}.}~\bibnamefont
  {Guazzelli}}\ and\ \bibinfo {author} {\bibfnamefont {O.}~\bibnamefont
  {Pouliquen}},\ }\href@noop {} {\bibfield  {journal} {\bibinfo  {journal} {J.
  Fluid Mech.}\ }\textbf {\bibinfo {volume} {852}},\ \bibinfo {pages} {P1}
  (\bibinfo {year} {2018})}\BibitemShut {NoStop}%
\bibitem [{\citenamefont {Ness}\ \emph {et~al.}(2022)\citenamefont {Ness},
  \citenamefont {Seto},\ and\ \citenamefont
  {Mari}}]{nessPhysicsDenseSuspensions2022}%
  \BibitemOpen
  \bibfield  {author} {\bibinfo {author} {\bibfnamefont {C.}~\bibnamefont
  {Ness}}, \bibinfo {author} {\bibfnamefont {R.}~\bibnamefont {Seto}},\ and\
  \bibinfo {author} {\bibfnamefont {R.}~\bibnamefont {Mari}},\ }\href
  {https://doi.org/10.1146/annurev-conmatphys-031620-105938} {\bibfield
  {journal} {\bibinfo  {journal} {Annu. Rev. Cond. Mat. Phys.}\ }\textbf
  {\bibinfo {volume} {13}},\ \bibinfo {pages} {97} (\bibinfo {year}
  {2022})}\BibitemShut {NoStop}%
\bibitem [{\citenamefont {Wyart}\ and\ \citenamefont
  {Cates}(2014)}]{wyartDiscontinuousShearThickening2014}%
  \BibitemOpen
  \bibfield  {author} {\bibinfo {author} {\bibfnamefont {M.}~\bibnamefont
  {Wyart}}\ and\ \bibinfo {author} {\bibfnamefont {M.~E.}\ \bibnamefont
  {Cates}},\ }\href {https://doi.org/10.1103/PhysRevLett.112.098302} {\bibfield
   {journal} {\bibinfo  {journal} {Phys. Rev. Lett.}\ }\textbf {\bibinfo
  {volume} {112}},\ \bibinfo {pages} {098302} (\bibinfo {year}
  {2014})}\BibitemShut {NoStop}%
\bibitem [{\citenamefont {Hsiao}\ \emph
  {et~al.}(2017{\natexlab{a}})\citenamefont {Hsiao}, \citenamefont
  {{Saha-Dalal}}, \citenamefont {Larson},\ and\ \citenamefont
  {Solomon}}]{hsiaoTranslationalRotationalDynamics2017}%
  \BibitemOpen
  \bibfield  {author} {\bibinfo {author} {\bibfnamefont {L.~C.}\ \bibnamefont
  {Hsiao}}, \bibinfo {author} {\bibfnamefont {I.}~\bibnamefont {{Saha-Dalal}}},
  \bibinfo {author} {\bibfnamefont {R.~G.}\ \bibnamefont {Larson}},\ and\
  \bibinfo {author} {\bibfnamefont {M.~J.}\ \bibnamefont {Solomon}},\ }\href
  {https://doi.org/10.1039/C7SM02115A} {\bibfield  {journal} {\bibinfo
  {journal} {Soft Matter}\ }\textbf {\bibinfo {volume} {13}},\ \bibinfo {pages}
  {9229} (\bibinfo {year} {2017}{\natexlab{a}})}\BibitemShut {NoStop}%
\bibitem [{\citenamefont {Ilhan}\ \emph {et~al.}(2022)\citenamefont {Ilhan},
  \citenamefont {Mugele},\ and\ \citenamefont
  {Duits}}]{ilhanRoughnessInducedRotational2022}%
  \BibitemOpen
  \bibfield  {author} {\bibinfo {author} {\bibfnamefont {B.}~\bibnamefont
  {Ilhan}}, \bibinfo {author} {\bibfnamefont {F.}~\bibnamefont {Mugele}},\ and\
  \bibinfo {author} {\bibfnamefont {M.~H.}\ \bibnamefont {Duits}},\ }\href
  {https://doi.org/10.1016/j.jcis.2021.08.212} {\bibfield  {journal} {\bibinfo
  {journal} {J. Colloid Interface Sci.}\ }\textbf {\bibinfo {volume} {607}},\
  \bibinfo {pages} {1709} (\bibinfo {year} {2022})}\BibitemShut {NoStop}%
\bibitem [{\citenamefont {Foss}\ and\ \citenamefont
  {Brady}(2000)}]{fossStructureDiffusionRheology2000}%
  \BibitemOpen
  \bibfield  {author} {\bibinfo {author} {\bibfnamefont {D.~R.}\ \bibnamefont
  {Foss}}\ and\ \bibinfo {author} {\bibfnamefont {J.~F.}\ \bibnamefont
  {Brady}},\ }\href {https://doi.org/10.1017/S0022112099007557} {\bibfield
  {journal} {\bibinfo  {journal} {J. Fluid Mech.}\ }\textbf {\bibinfo {volume}
  {407}},\ \bibinfo {pages} {167} (\bibinfo {year} {2000})}\BibitemShut
  {NoStop}%
\bibitem [{\citenamefont {Melrose}\ and\ \citenamefont
  {Ball}(2004)}]{melroseContinuousShearThickening2004}%
  \BibitemOpen
  \bibfield  {author} {\bibinfo {author} {\bibfnamefont {J.~R.}\ \bibnamefont
  {Melrose}}\ and\ \bibinfo {author} {\bibfnamefont {R.~C.}\ \bibnamefont
  {Ball}},\ }\href {https://doi.org/10.1122/1.1784783} {\bibfield  {journal}
  {\bibinfo  {journal} {J. Rheol.}\ }\textbf {\bibinfo {volume} {48}},\
  \bibinfo {pages} {937} (\bibinfo {year} {2004})}\BibitemShut {NoStop}%
\bibitem [{\citenamefont {Hsiao}\ \emph
  {et~al.}(2017{\natexlab{b}})\citenamefont {Hsiao}, \citenamefont {Jamali},
  \citenamefont {Glynos}, \citenamefont {Green}, \citenamefont {Larson},\ and\
  \citenamefont {Solomon}}]{hsiaoRheologicalStateDiagrams2017}%
  \BibitemOpen
  \bibfield  {author} {\bibinfo {author} {\bibfnamefont {L.~C.}\ \bibnamefont
  {Hsiao}}, \bibinfo {author} {\bibfnamefont {S.}~\bibnamefont {Jamali}},
  \bibinfo {author} {\bibfnamefont {E.}~\bibnamefont {Glynos}}, \bibinfo
  {author} {\bibfnamefont {P.~F.}\ \bibnamefont {Green}}, \bibinfo {author}
  {\bibfnamefont {R.~G.}\ \bibnamefont {Larson}},\ and\ \bibinfo {author}
  {\bibfnamefont {M.~J.}\ \bibnamefont {Solomon}},\ }\href
  {https://doi.org/10.1103/PhysRevLett.119.158001} {\bibfield  {journal}
  {\bibinfo  {journal} {Phys. Rev. Lett.}\ }\textbf {\bibinfo {volume} {119}},\
  \bibinfo {pages} {158001} (\bibinfo {year} {2017}{\natexlab{b}})}\BibitemShut
  {NoStop}%
\bibitem [{\citenamefont {Pradeep}\ \emph {et~al.}(2022)\citenamefont
  {Pradeep}, \citenamefont {Wessel},\ and\ \citenamefont
  {Hsiao}}]{pradeepHydrodynamicOriginSuspension2022}%
  \BibitemOpen
  \bibfield  {author} {\bibinfo {author} {\bibfnamefont {S.}~\bibnamefont
  {Pradeep}}, \bibinfo {author} {\bibfnamefont {A.}~\bibnamefont {Wessel}},\
  and\ \bibinfo {author} {\bibfnamefont {L.~C.}\ \bibnamefont {Hsiao}},\ }\href
  {https://doi.org/10.1122/8.0000424} {\bibfield  {journal} {\bibinfo
  {journal} {J. Rheol.}\ }\textbf {\bibinfo {volume} {66}},\ \bibinfo {pages}
  {895} (\bibinfo {year} {2022})}\BibitemShut {NoStop}%
\bibitem [{\citenamefont {Schroyen}\ \emph {et~al.}(2019)\citenamefont
  {Schroyen}, \citenamefont {Hsu}, \citenamefont {Isa}, \citenamefont
  {Van~Puyvelde},\ and\ \citenamefont
  {Vermant}}]{schroyenStressContributionsColloidal2019}%
  \BibitemOpen
  \bibfield  {author} {\bibinfo {author} {\bibfnamefont {B.}~\bibnamefont
  {Schroyen}}, \bibinfo {author} {\bibfnamefont {C.-P.}\ \bibnamefont {Hsu}},
  \bibinfo {author} {\bibfnamefont {L.}~\bibnamefont {Isa}}, \bibinfo {author}
  {\bibfnamefont {P.}~\bibnamefont {Van~Puyvelde}},\ and\ \bibinfo {author}
  {\bibfnamefont {J.}~\bibnamefont {Vermant}},\ }\href
  {https://doi.org/10.1103/PhysRevLett.122.218001} {\bibfield  {journal}
  {\bibinfo  {journal} {Phys. Rev. Lett.}\ }\textbf {\bibinfo {volume} {122}},\
  \bibinfo {pages} {218001} (\bibinfo {year} {2019})}\BibitemShut {NoStop}%
\bibitem [{\citenamefont {Yanagishima}\ \emph {et~al.}(2021)\citenamefont
  {Yanagishima}, \citenamefont {Liu}, \citenamefont {Tanaka},\ and\
  \citenamefont
  {Dullens}}]{yanagishimaParticleLevelVisualizationHydrodynamic2021}%
  \BibitemOpen
  \bibfield  {author} {\bibinfo {author} {\bibfnamefont {T.}~\bibnamefont
  {Yanagishima}}, \bibinfo {author} {\bibfnamefont {Y.}~\bibnamefont {Liu}},
  \bibinfo {author} {\bibfnamefont {H.}~\bibnamefont {Tanaka}},\ and\ \bibinfo
  {author} {\bibfnamefont {R.~P.~A.}\ \bibnamefont {Dullens}},\ }\href
  {https://doi.org/10.1103/PhysRevX.11.021056} {\bibfield  {journal} {\bibinfo
  {journal} {Phys. Rev. X}\ }\textbf {\bibinfo {volume} {11}},\ \bibinfo
  {pages} {021056} (\bibinfo {year} {2021})}\BibitemShut {NoStop}%
\bibitem [{\citenamefont {Hsu}\ \emph {et~al.}(2018)\citenamefont {Hsu},
  \citenamefont {Ramakrishna}, \citenamefont {Zanini}, \citenamefont
  {Spencer},\ and\ \citenamefont
  {Isa}}]{hsuRoughnessdependentTribologyEffects2018}%
  \BibitemOpen
  \bibfield  {author} {\bibinfo {author} {\bibfnamefont {C.-P.}\ \bibnamefont
  {Hsu}}, \bibinfo {author} {\bibfnamefont {S.~N.}\ \bibnamefont
  {Ramakrishna}}, \bibinfo {author} {\bibfnamefont {M.}~\bibnamefont {Zanini}},
  \bibinfo {author} {\bibfnamefont {N.~D.}\ \bibnamefont {Spencer}},\ and\
  \bibinfo {author} {\bibfnamefont {L.}~\bibnamefont {Isa}},\ }\href
  {https://doi.org/10.1073/pnas.1801066115} {\bibfield  {journal} {\bibinfo
  {journal} {Proc. Natl. Acad. Sci. U.S.A.}\ }\textbf {\bibinfo {volume}
  {115}},\ \bibinfo {pages} {5117} (\bibinfo {year} {2018})}\BibitemShut
  {NoStop}%
\bibitem [{\citenamefont {Seto}\ \emph {et~al.}(2013)\citenamefont {Seto},
  \citenamefont {Mari}, \citenamefont {Morris},\ and\ \citenamefont
  {Denn}}]{setoDiscontinuousShearThickening2013}%
  \BibitemOpen
  \bibfield  {author} {\bibinfo {author} {\bibfnamefont {R.}~\bibnamefont
  {Seto}}, \bibinfo {author} {\bibfnamefont {R.}~\bibnamefont {Mari}}, \bibinfo
  {author} {\bibfnamefont {J.~F.}\ \bibnamefont {Morris}},\ and\ \bibinfo
  {author} {\bibfnamefont {M.~M.}\ \bibnamefont {Denn}},\ }\href
  {https://doi.org/10.1103/PhysRevLett.111.218301} {\bibfield  {journal}
  {\bibinfo  {journal} {Phys. Rev. Lett.}\ }\textbf {\bibinfo {volume} {111}},\
  \bibinfo {pages} {218301} (\bibinfo {year} {2013})}\BibitemShut {NoStop}%
\bibitem [{\citenamefont {Fernandez}\ \emph {et~al.}(2013)\citenamefont
  {Fernandez}, \citenamefont {Mani}, \citenamefont {Rinaldi}, \citenamefont
  {Kadau}, \citenamefont {Mosquet}, \citenamefont {{Lombois-Burger}},
  \citenamefont {{Cayer-Barrioz}}, \citenamefont {Herrmann}, \citenamefont
  {Spencer},\ and\ \citenamefont
  {Isa}}]{fernandezMicroscopicMechanismShear2013}%
  \BibitemOpen
  \bibfield  {author} {\bibinfo {author} {\bibfnamefont {N.}~\bibnamefont
  {Fernandez}}, \bibinfo {author} {\bibfnamefont {R.}~\bibnamefont {Mani}},
  \bibinfo {author} {\bibfnamefont {D.}~\bibnamefont {Rinaldi}}, \bibinfo
  {author} {\bibfnamefont {D.}~\bibnamefont {Kadau}}, \bibinfo {author}
  {\bibfnamefont {M.}~\bibnamefont {Mosquet}}, \bibinfo {author} {\bibfnamefont
  {H.}~\bibnamefont {{Lombois-Burger}}}, \bibinfo {author} {\bibfnamefont
  {J.}~\bibnamefont {{Cayer-Barrioz}}}, \bibinfo {author} {\bibfnamefont
  {H.~J.}\ \bibnamefont {Herrmann}}, \bibinfo {author} {\bibfnamefont {N.~D.}\
  \bibnamefont {Spencer}},\ and\ \bibinfo {author} {\bibfnamefont
  {L.}~\bibnamefont {Isa}},\ }\href
  {https://doi.org/10.1103/PhysRevLett.111.108301} {\bibfield  {journal}
  {\bibinfo  {journal} {Phys. Rev. Lett.}\ }\textbf {\bibinfo {volume} {111}},\
  \bibinfo {pages} {108301} (\bibinfo {year} {2013})}\BibitemShut {NoStop}%
\bibitem [{\citenamefont {Mari}\ \emph {et~al.}(2014)\citenamefont {Mari},
  \citenamefont {Seto}, \citenamefont {Morris},\ and\ \citenamefont
  {Denn}}]{mariShearThickeningFrictionless2014}%
  \BibitemOpen
  \bibfield  {author} {\bibinfo {author} {\bibfnamefont {R.}~\bibnamefont
  {Mari}}, \bibinfo {author} {\bibfnamefont {R.}~\bibnamefont {Seto}}, \bibinfo
  {author} {\bibfnamefont {J.~F.}\ \bibnamefont {Morris}},\ and\ \bibinfo
  {author} {\bibfnamefont {M.~M.}\ \bibnamefont {Denn}},\ }\href
  {https://doi.org/10.1122/1.4890747} {\bibfield  {journal} {\bibinfo
  {journal} {J. Rheol.}\ }\textbf {\bibinfo {volume} {58}},\ \bibinfo {pages}
  {1693} (\bibinfo {year} {2014})}\BibitemShut {NoStop}%
\bibitem [{\citenamefont {Townsend}\ and\ \citenamefont
  {Wilson}(2017)}]{townsendFrictionalShearThickening2017}%
  \BibitemOpen
  \bibfield  {author} {\bibinfo {author} {\bibfnamefont {A.~K.}\ \bibnamefont
  {Townsend}}\ and\ \bibinfo {author} {\bibfnamefont {H.~J.}\ \bibnamefont
  {Wilson}},\ }\href {https://doi.org/10.1063/1.4989929} {\bibfield  {journal}
  {\bibinfo  {journal} {Phys. Fluids}\ }\textbf {\bibinfo {volume} {29}},\
  \bibinfo {pages} {121607} (\bibinfo {year} {2017})}\BibitemShut {NoStop}%
\bibitem [{\citenamefont {Jamali}\ and\ \citenamefont
  {Brady}(2019)}]{jamaliAlternativeFrictionalModel2019a}%
  \BibitemOpen
  \bibfield  {author} {\bibinfo {author} {\bibfnamefont {S.}~\bibnamefont
  {Jamali}}\ and\ \bibinfo {author} {\bibfnamefont {J.~F.}\ \bibnamefont
  {Brady}},\ }\href {https://doi.org/10.1103/PhysRevLett.123.138002} {\bibfield
   {journal} {\bibinfo  {journal} {Phys. Rev. Lett.}\ }\textbf {\bibinfo
  {volume} {123}},\ \bibinfo {pages} {138002} (\bibinfo {year}
  {2019})}\BibitemShut {NoStop}%
\bibitem [{\citenamefont {Cunha}\ and\ \citenamefont
  {Hinch}(1996)}]{cunhaShearinducedDispersionDilute1996}%
  \BibitemOpen
  \bibfield  {author} {\bibinfo {author} {\bibfnamefont {F.~R.~D.}\
  \bibnamefont {Cunha}}\ and\ \bibinfo {author} {\bibfnamefont {E.~J.}\
  \bibnamefont {Hinch}},\ }\href {https://doi.org/10.1017/S0022112096001619}
  {\bibfield  {journal} {\bibinfo  {journal} {J. Fluid Mech.}\ }\textbf
  {\bibinfo {volume} {309}},\ \bibinfo {pages} {211} (\bibinfo {year}
  {1996})}\BibitemShut {NoStop}%
\bibitem [{\citenamefont {Blanc}\ \emph {et~al.}(2011)\citenamefont {Blanc},
  \citenamefont {Peters},\ and\ \citenamefont
  {Lemaire}}]{blancExperimentalSignaturePair2011}%
  \BibitemOpen
  \bibfield  {author} {\bibinfo {author} {\bibfnamefont {F.}~\bibnamefont
  {Blanc}}, \bibinfo {author} {\bibfnamefont {F.}~\bibnamefont {Peters}},\ and\
  \bibinfo {author} {\bibfnamefont {E.}~\bibnamefont {Lemaire}},\ }\href
  {https://doi.org/10.1103/PhysRevLett.107.208302} {\bibfield  {journal}
  {\bibinfo  {journal} {Phys. Rev. Lett.}\ }\textbf {\bibinfo {volume} {107}},\
  \bibinfo {pages} {208302} (\bibinfo {year} {2011})}\BibitemShut {NoStop}%
\bibitem [{\citenamefont {Wang}\ \emph {et~al.}(2020)\citenamefont {Wang},
  \citenamefont {Jamali},\ and\ \citenamefont
  {Brady}}]{wangHydrodynamicModelDiscontinuous2020a}%
  \BibitemOpen
  \bibfield  {author} {\bibinfo {author} {\bibfnamefont {M.}~\bibnamefont
  {Wang}}, \bibinfo {author} {\bibfnamefont {S.}~\bibnamefont {Jamali}},\ and\
  \bibinfo {author} {\bibfnamefont {J.~F.}\ \bibnamefont {Brady}},\ }\href
  {https://doi.org/10.1122/1.5134036} {\bibfield  {journal} {\bibinfo
  {journal} {J. Rheol.}\ }\textbf {\bibinfo {volume} {64}},\ \bibinfo {pages}
  {379} (\bibinfo {year} {2020})}\BibitemShut {NoStop}%
\bibitem [{\citenamefont {Kurzthaler}\ \emph {et~al.}(2020)\citenamefont
  {Kurzthaler}, \citenamefont {Zhu}, \citenamefont {Pahlavan},\ and\
  \citenamefont {Stone}}]{kurzthalerParticleMotionNearby2020}%
  \BibitemOpen
  \bibfield  {author} {\bibinfo {author} {\bibfnamefont {C.}~\bibnamefont
  {Kurzthaler}}, \bibinfo {author} {\bibfnamefont {L.}~\bibnamefont {Zhu}},
  \bibinfo {author} {\bibfnamefont {A.~A.}\ \bibnamefont {Pahlavan}},\ and\
  \bibinfo {author} {\bibfnamefont {H.~A.}\ \bibnamefont {Stone}},\ }\href
  {https://doi.org/10.1103/PhysRevFluids.5.082101} {\bibfield  {journal}
  {\bibinfo  {journal} {Phys. Rev. Fluids}\ }\textbf {\bibinfo {volume} {5}},\
  \bibinfo {pages} {082101} (\bibinfo {year} {2020})}\BibitemShut {NoStop}%
\bibitem [{\citenamefont {Chase}\ \emph {et~al.}(2022)\citenamefont {Chase},
  \citenamefont {Kurzthaler},\ and\ \citenamefont
  {Stone}}]{chaseHydrodynamicallyInducedHelical2022}%
  \BibitemOpen
  \bibfield  {author} {\bibinfo {author} {\bibfnamefont {D.~L.}\ \bibnamefont
  {Chase}}, \bibinfo {author} {\bibfnamefont {C.}~\bibnamefont {Kurzthaler}},\
  and\ \bibinfo {author} {\bibfnamefont {H.~A.}\ \bibnamefont {Stone}},\ }\href
  {https://doi.org/10.1073/pnas.2202082119} {\bibfield  {journal} {\bibinfo
  {journal} {Proc. Natl. Acad. Sci. U.S.A.}\ }\textbf {\bibinfo {volume}
  {119}},\ \bibinfo {pages} {e2202082119} (\bibinfo {year} {2022})}\BibitemShut
  {NoStop}%
\bibitem [{\citenamefont {Yariv}\ \emph
  {et~al.}(2024{\natexlab{a}})\citenamefont {Yariv}, \citenamefont
  {Brand{\~a}o}, \citenamefont {Wood}, \citenamefont {Szafraniec},
  \citenamefont {Higgins}, \citenamefont {Bazazi}, \citenamefont {Pearce},\
  and\ \citenamefont {Stone}}]{yarivHydrodynamicInteractionsRough2024}%
  \BibitemOpen
  \bibfield  {author} {\bibinfo {author} {\bibfnamefont {E.}~\bibnamefont
  {Yariv}}, \bibinfo {author} {\bibfnamefont {R.}~\bibnamefont {Brand{\~a}o}},
  \bibinfo {author} {\bibfnamefont {D.~K.}\ \bibnamefont {Wood}}, \bibinfo
  {author} {\bibfnamefont {H.}~\bibnamefont {Szafraniec}}, \bibinfo {author}
  {\bibfnamefont {J.~M.}\ \bibnamefont {Higgins}}, \bibinfo {author}
  {\bibfnamefont {P.}~\bibnamefont {Bazazi}}, \bibinfo {author} {\bibfnamefont
  {P.}~\bibnamefont {Pearce}},\ and\ \bibinfo {author} {\bibfnamefont {H.~A.}\
  \bibnamefont {Stone}},\ }\href
  {https://doi.org/10.1103/PhysRevFluids.9.L032301} {\bibfield  {journal}
  {\bibinfo  {journal} {Phys. Rev. Fluids}\ }\textbf {\bibinfo {volume} {9}},\
  \bibinfo {pages} {L032301} (\bibinfo {year}
  {2024}{\natexlab{a}})}\BibitemShut {NoStop}%
\bibitem [{SI()}]{SI}%
  \BibitemOpen
  \href@noop {} {\bibinfo {title} {See {S}upplemental {M}aterial for additional
  theoretical and numerical details.}}\BibitemShut {Stop}%
\bibitem [{\citenamefont {Ockendon}\ and\ \citenamefont
  {Ockendon}(1995)}]{ockendon1995viscous}%
  \BibitemOpen
  \bibfield  {author} {\bibinfo {author} {\bibfnamefont {H.}~\bibnamefont
  {Ockendon}}\ and\ \bibinfo {author} {\bibfnamefont {J.~R.}\ \bibnamefont
  {Ockendon}},\ }\href@noop {} {\emph {\bibinfo {title} {Viscous Flow}}}\
  (\bibinfo  {publisher} {Cambridge University Press},\ \bibinfo {address} {New
  York, USA},\ \bibinfo {year} {1995})\BibitemShut {NoStop}%
\bibitem [{\citenamefont {Press}(2007)}]{press2007numerical}%
  \BibitemOpen
  \bibfield  {author} {\bibinfo {author} {\bibfnamefont {W.~H.}\ \bibnamefont
  {Press}},\ }\href@noop {} {\emph {\bibinfo {title} {Numerical Recipes 3rd
  Edition: The Art of Scientific Computing}}}\ (\bibinfo  {publisher}
  {Cambridge University Press},\ \bibinfo {year} {2007})\BibitemShut {NoStop}%
\bibitem [{\citenamefont {Jeffrey}\ and\ \citenamefont
  {Onishi}(1984)}]{jeffreyCalculationResistanceMobility1984}%
  \BibitemOpen
  \bibfield  {author} {\bibinfo {author} {\bibfnamefont {D.~J.}\ \bibnamefont
  {Jeffrey}}\ and\ \bibinfo {author} {\bibfnamefont {Y.}~\bibnamefont
  {Onishi}},\ }\href {https://doi.org/10.1017/S0022112084000355} {\bibfield
  {journal} {\bibinfo  {journal} {J. Fluid Mech.}\ }\textbf {\bibinfo {volume}
  {139}},\ \bibinfo {pages} {261} (\bibinfo {year} {1984})}\BibitemShut
  {NoStop}%
\bibitem [{\citenamefont {Cox}(1974)}]{cox_suspended_particles_1974}%
  \BibitemOpen
  \bibfield  {author} {\bibinfo {author} {\bibfnamefont {R.}~\bibnamefont
  {Cox}},\ }\href
  {https://doi.org/https://doi.org/10.1016/0301-9322(74)90019-6} {\bibfield
  {journal} {\bibinfo  {journal} {Int. J. Multiphase Flow}\ }\textbf {\bibinfo
  {volume} {1}},\ \bibinfo {pages} {343} (\bibinfo {year} {1974})}\BibitemShut
  {NoStop}%
\bibitem [{\citenamefont {Yariv}\ \emph
  {et~al.}(2024{\natexlab{b}})\citenamefont {Yariv}, \citenamefont {Brand\~ao},
  \citenamefont {Wood}, \citenamefont {Szafraniec}, \citenamefont {Higgins},
  \citenamefont {Bazazi}, \citenamefont {Pearce},\ and\ \citenamefont
  {Stone}}]{yariv_rough_surfaces_2024}%
  \BibitemOpen
  \bibfield  {author} {\bibinfo {author} {\bibfnamefont {E.}~\bibnamefont
  {Yariv}}, \bibinfo {author} {\bibfnamefont {R.}~\bibnamefont {Brand\~ao}},
  \bibinfo {author} {\bibfnamefont {D.~K.}\ \bibnamefont {Wood}}, \bibinfo
  {author} {\bibfnamefont {H.}~\bibnamefont {Szafraniec}}, \bibinfo {author}
  {\bibfnamefont {J.~M.}\ \bibnamefont {Higgins}}, \bibinfo {author}
  {\bibfnamefont {P.}~\bibnamefont {Bazazi}}, \bibinfo {author} {\bibfnamefont
  {P.}~\bibnamefont {Pearce}},\ and\ \bibinfo {author} {\bibfnamefont {H.~A.}\
  \bibnamefont {Stone}},\ }\href
  {https://doi.org/10.1103/PhysRevFluids.9.L032301} {\bibfield  {journal}
  {\bibinfo  {journal} {Phys. Rev. Fluids}\ }\textbf {\bibinfo {volume} {9}},\
  \bibinfo {pages} {L032301} (\bibinfo {year}
  {2024}{\natexlab{b}})}\BibitemShut {NoStop}%
\end{thebibliography}

%

\end{document}